\documentclass[useAMS,usenatbib]{mn2e}

\usepackage{graphicx}
\usepackage[bookmarks=false, colorlinks=true, citecolor=blue, linkcolor=blue, urlcolor=blue]{hyperref}

\def\aj{AJ}%
\def\araa{ARA\&A}%
\def\apj{ApJ}%
\def\apjl{ApJ}%
\def\apjs{ApJS}%
\def\aap{A\&A}%
\def\mnras{MNRAS}%
\DeclareRobustCommand{\ion}[2]{%
\relax\ifmmode
\ifx\testbx\f@series
{\mathbf{#1\,\mathsc{#2}}}\else
{\mathrm{#1\,\mathsc{#2}}}\fi
\else\textup{#1\,{\mdseries\textsc{#2}}}%
\fi}

\newcommand{\Spitzer}{\textit{Spitzer}}
\def\micron{\hbox{\,$\mu$m}}
\newcommand{\Lsun}{\hbox{$L_{\rm \odot}$}}
\newcommand{\Msun}{\hbox{$M_{\rm \odot}$}}

\newcommand{\degree}{\ensuremath{^\circ}}
\newcommand\nodata{ ~$\cdots$~ }

\title[Mid-IR observations of NGC~1614]{Sub-arcsec mid-IR observations of NGC~1614:
Nuclear star-formation or an intrinsically X-ray weak AGN?}
\author[Pereira-Santaella et al.]{\parbox{\textwidth}{M. Pereira-Santaella$^{1,2,}$\thanks{E-mail:
mpereira@cab.inta-csic.es}, L. Colina$^{1,2}$, A. Alonso-Herrero$^{3}$, A. Usero$^{4}$, T. D\'iaz-Santos$^{5}$, S. Garc\'ia-Burillo$^{4}$, A. Alberdi$^{6}$, O. Gonzalez-Martin$^7$, R. Herrero-Illana$^{6}$, M. Imanishi $^8$, N.~A. Levenson$^9$, M.~A. P\'erez-Torres$^{6,10}$, C. Ramos Almeida$^{11}$
 }
\vspace{0.4cm} \\
$^{1}$Centro de Astrobiolog\'ia (CSIC/INTA), Ctra de Torrej\'on a Ajalvir, km 4, 28850, Torrej\'on de Ardoz, Madrid, Spain\\
$^{2}$ASTRO-UAM, UAM, Unidad Asociada CSIC\\
$^{3}$Instituto de F\'isica de Cantabria, CSIC-Universidad de Cantabria, 39005 Santander, Spain\\
$^{4}$Observatorio Astron\'omico Nacional (OAN-IGN)-Observatorio de Madrid, Alfonso XII, 3, 28014, Madrid, Spain \\
$^{5}$N\'ucleo de Astronom\'ia de la Facultad de Ingenier\'ia, Universidad Diego Portales, Av. Ej\'ercito Libertador 441, Santiago, Chile\\
$^{6}$Instituto de Astrof\'isica de Andaluc\'ia, Glorieta de las Astronom\'ia, s/n, 18008 Granada, Spain\\
$^{7}$Centro de Radioastronom\'ia y Astrof\'isica (CRyA-UNAM), 3-72 (Xangari), 8701, Morelia, Mexico\\
$^{8}$Subaru Telescope, 650 North A'ohoku Place, Hilo, Hawaii, 96720, U.S.A.\\
$^{9}$Gemini Observatory, Casilla 603, La Serena, Chile\\
$^{10}$Centro de Estudios de la F\'isica del Cosmos de Arag\'on, 44001 Teruel, Spain\\
$^{11}$Instituto de Astrof\'isica de Canarias, V\'ia L\'actea s/n, 38205 La Laguna, Tenerife, Spain
}
\begin{document}

\date{}

\maketitle

\label{firstpage}
\begin{abstract}
We present new mid-infrared $N$-band spectroscopy and $Q$-band photometry of the local luminous infrared  galaxy NGC~1614, one of the most extreme nearby starbursts. We analyze the mid-IR properties of the nucleus (central 150\,pc) and four regions of the bright circumnuclear (diameter$\sim 600$\,pc) star-forming (SF) ring of this object. The nucleus differs from the circumnuclear SF ring by having a {strong 8--12\micron\ continuum} (low 11.3\micron\ PAH equivalent width).
These characteristics, together with the nuclear X-ray and sub-mm properties, can be explained by an X-ray weak active galactic nucleus (AGN), or by peculiar SF with a short molecular gas depletion time and producing an enhanced radiation field density. In either case, the nuclear luminosity ($L_{\rm IR}<$6$\times$10$^{43}$\,erg\,s$^{-1}$) is only $<$5\,\%\ of the total bolometric luminosity of NGC~1614. So this possible AGN does not dominate the energy output in this object.
We also compare three star-formation rate (SFR) tracers (Pa$\alpha$, 11.3\micron\ PAH, and 24\micron\ emissions) at 150\,pc\ scales {in the circumnuclear ring}. In general, we find that the SFR is underestimated (overestimated) by a factor of 2--4 (2--3) using the 11.3\micron\ PAH (24\micron) emission with respect to the extinction corrected Pa$\alpha$ SFR. The former can be explained because we do not include diffuse PAH emission in our measurements, while the latter might indicate that the dust temperature is particularly warmer in the central regions of NGC~1614.
\end{abstract}


\section{Introduction}\label{s:intro}

Ultra-luminous and luminous infrared galaxies (U\slash LIRGs) are objects with infrared (IR) luminosities ($ L_{\rm IR}$) between 10$^{11}$ and 10$^{12}$\,\Lsun\ (LIRGs) and $>$10$^{12}$ \Lsun\ (ULIRGs). Locally, objects with such high IR luminosities are unusual. However, between $z\sim 1$ and 2, galaxies in the LIRG and ULIRG luminosity ranges dominate the star-formation rate (SFR) density of the Universe \citep{PerezGonzalez2005, LeFloch2005, Caputi2007, Magnelli2011}.
Therefore, the study at high-angular resolution of local LIRGs provides a unique insight into extreme SF environments similar to those of high-$z$ galaxies near the SFR density peak of the Universe \citep{Madau2014}.

NGC~1614 (Mrk~617) is the second most luminous galaxy within 75\,Mpc ($\log L_{\rm IR}=11.6$; \citealt{SandersRBGS}) and according to optical spectroscopy its nuclear activity is classified as composite \citep{Yuan2010}. 
It is an advanced minor merger (3:1--5:1 mass ratio; \citealt{Vaisanen2012}) located at 64\,Mpc (310\,pc\,arcsec$^{-1}$) with long tidal tails. Its bolometric luminosity is dominated by a strong starburst in the central kpc \citep{AAH01, Imanishi2010}, and, so far, there is no clear evidence of an active galactic nucleus (AGN) in NGC~1614 \citep{Herrero-Illana2014}.

The central kpc of NGC~1614 contains a compact nucleus (45-80\,pc), which dominates the near-IR continuum emission, and a bright circumnuclear SF ring (diameter$\sim600$\,pc), which is predominant in Pa$\alpha$ \citep{AAH01} and other SF indicators like the polycyclic aromatic hydrocarbon (PAH) emission \citep{DiazSantos2008,Vaisanen2012}, cold molecular gas \citep{Konig2013, Sliwa2014, Xu2015}, and radio continuum \citep{Olsson2010, Herrero-Illana2014}. In addition, \citet{GarciaBurillo2015} found a massive cold molecular gas outflow (3$\times$10$^7$\,$M_\odot$; $\dot{M}_{\rm out}\sim$40\,$M_{\odot}$\,yr$^{-1}$) which can be powered by the SF in the ring.

A bright obscured AGN is discarded by X-ray observations \citep{Pereira2011, Herrero-Illana2014}. However, previous mid-IR $N$-band imaging of NGC~1614 showed that the compact nucleus has a relatively high surface brightness \citep{Soifer2001, DiazSantos2008, Siebenmorgen2008}. Therefore, these observations suggest an enhanced mid-IR luminosity to SFR (as inferred from the observed Pa$\alpha$ luminosity) ratio in the nucleus \citep{DiazSantos2008}, which might indicate the presence of an active nucleus. However, without high angular resolution spectroscopy no detailed studies were possible.

\begin{figure*}
\centering
\includegraphics[width=\textwidth]{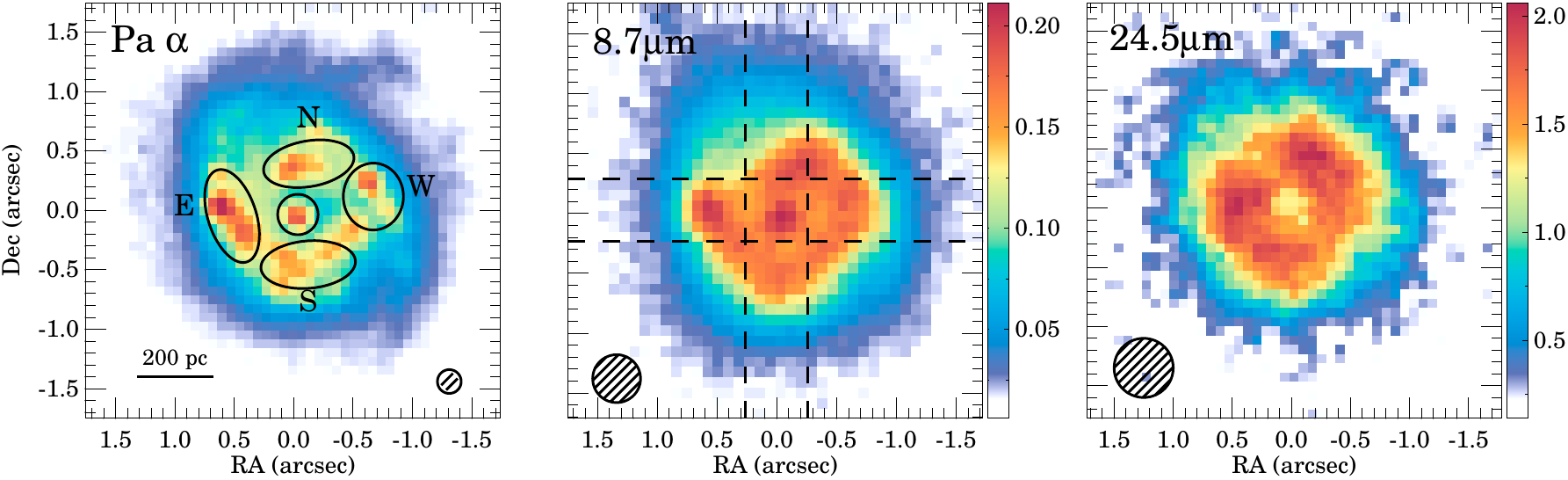}
\caption{\textit{HST}\slash NICMOS continuum subtracted Pa$\alpha$ (left; see also \citealt{AAH06s}), T-ReCS 8.7\micron\ (middle; see also \citealt{DiazSantos2008}), and CanariCam 24.5\micron\ (right) images of the nuclear regions of NGC~1614. The 24.5\micron\ image was smoothed using a 1.5\,pixel Gaussian.
The position and width of the two slit orientations are plotted in the middle panel (dashed lines). The hatched circles indicate the angular resolution of the images (FWHM). For the 8.7 and 24.5\micron\ images the color scale is in Jy\,arcsec$^{-2}$ units. The locations of the regions used for the spatial decomposition are indicated in the left panel.
\label{fig:maps}}
\end{figure*}

In this paper we present the first high-angular resolution ($\sim$0\farcs5 ) $N$-band (7.5--13\micron) spectroscopy of the nucleus and surrounding star-forming ring of NGC~1614, as well as $Q$-band 24.5\micron\ imaging using CanariCam on the 10.4\,m Gran Telescopio CANARIAS (GTC). First, we describe the new observations in Section \ref{s:data}. The extraction of the spectra and photometry, and a simple two component modeling are presented in Section \ref{s:analysis}. We explore the AGN or SF nature of the nucleus in Section \ref{ss:agn_vs_sf}, and, in Section \ref{ss:sfr_tracers}, the reliability of several SFR tracers at 150\,pc scales is discussed. The main conclusions are presented in Section \ref{s:conclusions}.

Throughout this paper we assume the following cosmology $H_{\rm 0} = 70$\,km\,s$^{-1}$\,Mpc$^{-1}$, $\Omega_{\rm m}=0.3$, and $\Omega_{\rm \Lambda}=0.7$ and the \citet{Kroupa2001} IMF.

\section{Observations and Data Reduction}\label{s:data}

\subsection{Mid-IR Imaging}

We obtained $Q$-band diffraction limited (0\farcs5) images of NGC~1614 using the Q8 filter ($\lambda_{\rm c}=24.5$\micron, width at 50\% cut-on/off of $\Delta\lambda = 0.8$\micron) of CanariCam (CC; \citealt{Telesco2003CC}) on the 10.4\,m GTC during December 2nd 2014.
These observations are part of the ESO/GTC large program 182.B-2005 (PI Alonso-Herrero).
The plate scale of CC is 0\farcs08\slash pixel and its field of view is 26\arcsec$\times$19\arcsec, so it covers the central 6\,kpc of NGC~1614.

Three exposures were taken with an on-source integration of 400\,s each. To reduce the data we used the \textsc{redcan} pipeline \citep{GonzalezMartin2013RedCan}. It performs the flat-fielding, stacking, and flux calibration of the individual exposures. The three reduced images were then combined after correcting the different background levels (right panel of Figure \ref{fig:maps}). For the flux calibration the standard star HD~28749 was observed. It is relatively weak at 24.5\micron\ (1.2\,Jy; \citealt{Cohen1999}) so the absolute calibration error of our $Q$-band observations is $\sim$20\,\%.

To check the flux calibration we also compared the integrated flux of NGC~1614 in our 24.5\micron\ image (6.0$\pm$0.9\,Jy) with the \Spitzer\slash MIPS 24\micron\ flux (5.7$\pm$0.3\,Jy; \citealt{Pereira2015not}). Both values are in good agreement.

In addition, $N$-band imaging of this galaxy was previously obtained using Gemini/T-ReCS in the Si2 filter ($\lambda_{\rm c}=8.7$\micron, $\Delta\lambda = 0.8$\micron). This image was published by \citet{DiazSantos2008} and it is shown in the middle panel of Figure \ref{fig:maps}. The angular resolution of this observation estimated from the calibration star image is 0\farcs4.

\subsection{Mid-IR Spectroscopy}

We obtained $N$-band spectroscopy (7.5--13\micron) of NGC~1614 with GTC\slash CanariCam on September 8th 2013 and January 5th 2014.
The low spectral resolution ($R\sim175$) grating was used. These observations are also part of the ESO/GTC large program 182.B-2005.
The nucleus of NGC~1614 was observed with a slit of 0\farcs52 width using two perpendicular orientations (position angles 0 and 90\degree). The approximate location of the slits is overplotted in the middle panel of Figure \ref{fig:maps}. The on-source integration time for each of the slit orientations was 1200\,s.

The standard star HD~28749 was observed in spectroscopy mode to provide the absolute flux calibration and telluric correction. From the two-dimensional spectrum of the standard star we derive that the angular resolution, $\sim$0\farcs5, is approximately constant with the wavelength both nights. That is, the spectroscopy was not obtained in diffraction limited conditions.

The data were reduced using the \textsc{redcan} pipeline. Flat-fielding, stacking, wavelength calibration, and flux calibration of the exposures are performed by this software. The spectra were extracted using a custom procedure (see Section \ref{ss:image_modeling}) instead of the default \textsc{redcan} extraction.

\section{Analysis and Results}\label{s:analysis}

\subsection{Image modeling and spectral extraction}\label{ss:image_modeling}
The \textit{HST}/NICMOS Pa$\alpha$ image of NGC~1614 (\citealt{AAH01}, see Fig 1) revealed that the angular separation between the nucleus and the star-forming ring is 0\farcs5--0\farcs7, which is comparable to the angular resolution of the CC Q-band imaging and N-band spectroscopy. Therefore, to disentangle the emission produced by the different regions we modeled the mid-IR image with the highest resolution (i.e., the 8.7\micron\ T-ReCS image) with \textsc{galfit} \citep{Peng2010GALFIT}.
\citet{Imanishi2011} published $Q$-band imaging of NGC~1614 at 17.7\micron. In this image, the emissions from the SF ring and the nucleus are not as clearly separated as in the CC 24.5\micron\ image, probably due to the slightly worse angular resolution (0\farcs7; \citealt{Asmus2014}).

\begin{figure}
\centering
\includegraphics[width=0.5\textwidth]{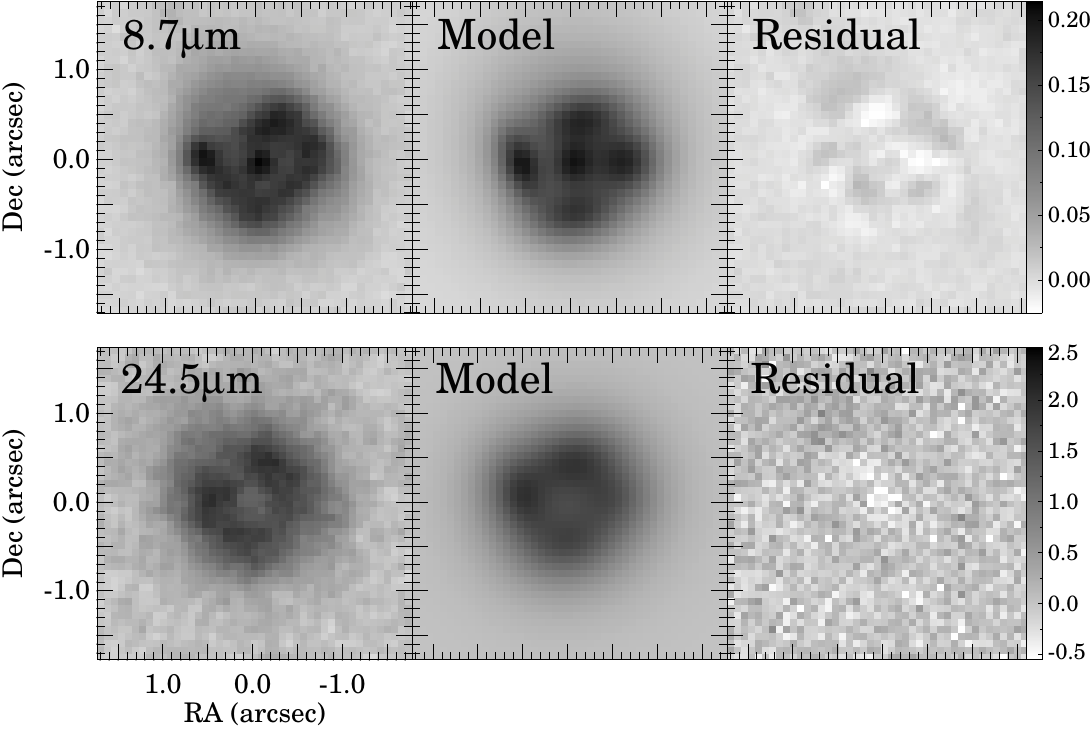}
\caption{\textsc{galfit} models of the T-ReCS 8.7\micron\ (top panels), and CanariCam 24.5\micron\ (bottom panels) observations of the nuclear regions of NGC~1614. The observed image, the best model and the residuals are shown in the left, middle, and right panels, respectively. The color scale is in Jy\,arcsec$^{-2}$ units. 
\label{fig:galfit}}
\end{figure}

We used six Gaussian spatial components (nucleus, north, south, east, west, and diffuse) convolved with the PSF, to reproduce the 8.7\micron\ image (see Figure \ref{fig:galfit}). These components are motivated by the Pa$\alpha$ morphology (Figure \ref{fig:maps}) and is the minimum number of components needed to reproduce the mid-IR images.
The position and full width half maximum (FWHM) of these components are listed in Table \ref{tab:components}. According to this decomposition, the ring is located $\sim$0\farcs6 away from the nucleus (190\,pc) and has a FWHM of $\sim$0\farcs5--0\farcs7 (160--220\,pc). The residuals of the model are less than 20\,\%\ (Figure \ref{fig:galfit}).

To extract the fluxes from the CC Q-band image, we used \textsc{galfit} fixing the relative positions and widths of these components, but allowing their intensities to vary (see Figure \ref{fig:galfit} and Table \ref{tab:fluxes}).

\begin{table}
\centering
\caption{Spatial decomposition nuclear region and circumnuclear ring of star formation of NGC~1614\label{tab:components}}
\begin{tabular}{@{}lcccccccccc@{}}
\hline
Region & $d^a$  & FWHM$^b$   \\
& (arcsec) & (arcsec) \\
\hline
Nucleus & \nodata & 0.21 \\
Diffuse & \nodata & 2.4 \\
N  & 0.52 & 0.73 \\
S  & 0.59 & 0.72 \\
E  & 0.60 & 0.54 \\
W  & 0.61 & 0.45 \\
\hline
\end{tabular}

\medskip
\raggedright \textbf{Notes:} $^{(a)}$ Angular distance between the nucleus and the component. $^{(b)}$ Deconvolved FWHM.
\end{table}

\begin{figure}
\centering
\includegraphics[width=0.46\textwidth]{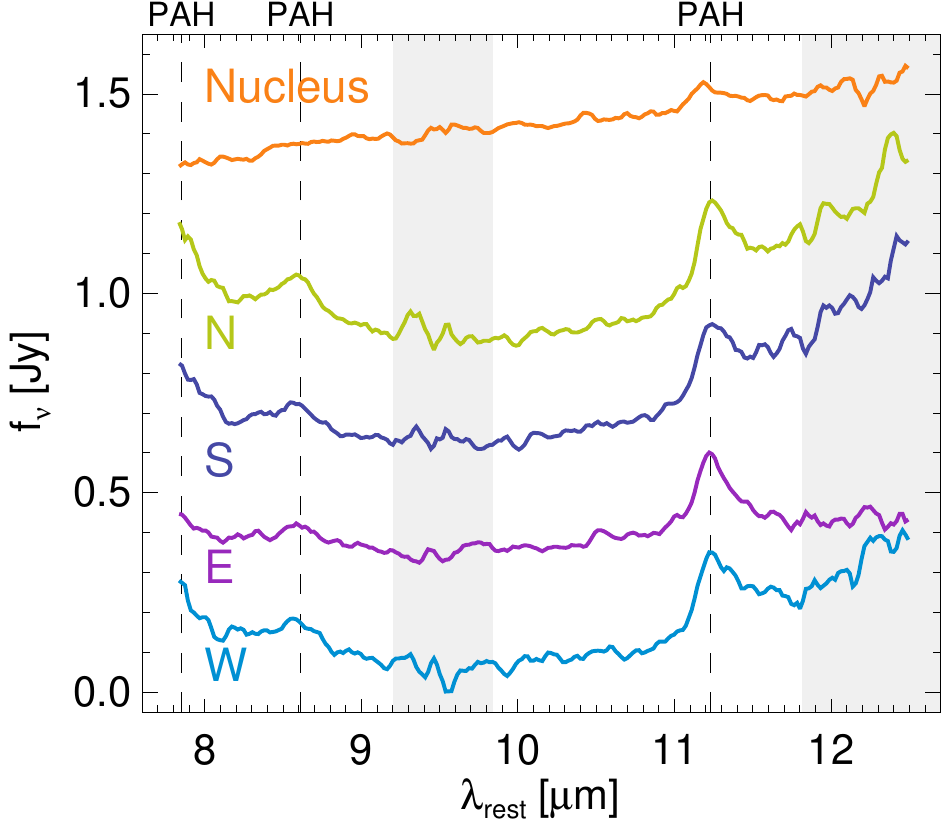}
\caption{Mid-IR CanariCam spectra of the nucleus and different regions in the star-forming ring. The nuclear (orange), north (green), south (dark blue), and east (purple) spectra are shifted by 1.3, 0.8, 0.55, and 0.3\,Jy, respectively. 
The vertical lines mark the wavelength of the 7.7, 8.6, and 11.3\micron\ PAH features (dashed line). The shaded gray regions mark low-atmospheric transmission spectral ranges.
\label{fig:cc_spec}}
\end{figure}

Similarly, we used this information to extract the CC N-band spectra. For each wavelength we generated a synthetic image taking into account the CC N-band PSF, and then we simulated the two slit orientations (P.A. 0 and 90\degree) to obtain the one-dimensional spatial profiles. We varied the intensities of the different regions to reproduce simultaneously the observed N-S and E-W profiles at each wavelength. The resulting spectra are plotted in Figure \ref{fig:cc_spec}. The fluxes at 10 and 12\micron, and the 11.3\micron\ PAH flux and equivalent width (EW) are listed in Table \ref{tab:fluxes}.

\begin{table*}
\caption{Spatially resolved measurements of NGC~1614.\label{tab:fluxes}}
\begin{tabular}{@{}lcccccccccc@{}}
\hline
Region &  f$_\nu$(10\micron)$^a$ & f$_\nu$(12\micron)$^a$ & f$_\nu$(24.1\micron)$^b$ & 11.3\micron\ PAH$^c$ & ${EW_{\rm 11.3\micron}}^d$  & f$_\nu$(Pa$\alpha$)$^e$ \\
 & (mJy) & (mJy) & (Jy) & (10$^{-13}$\,erg\,cm$^{-2}$\,s$^{-1}$) & (10$^{-3}$\micron) & (mJy) \\
\hline
Nucleus  & 120 $\pm$ 8 & 210 $\pm$ 20 & $<$0.5 & 3.2 $\pm$ 0.6 & 79 $\pm$ 8 & 2.2 \\
N & 91 $\pm$ 10 & 390 $\pm$ 30 & 1.7 $\pm$ 0.6 & 12.7 $\pm$ 0.2 & 220 $\pm$ 5 & 5.2 \\
S & 81 $\pm$ 10 & 390 $\pm$ 50 & 1.6 $\pm$ 0.4 & 9.3 $\pm$ 0.5 & 170 $\pm$ 3 & 5.0 \\
E & 61 $\pm$ 6 & 130 $\pm$ 10 & 2.0 $\pm$ 0.7 & 13.2 $\pm$ 0.3 & 500 $\pm$ 20 & 5.5 \\
W & 73 $\pm$ 20 & 280 $\pm$ 30 & 0.6 $\pm$ 0.3 & 14.2 $\pm$ 0.4 & 350 $\pm$ 10 & 4.6 \\
\hline
\end{tabular}

\medskip
\raggedright \textbf{Notes:} 3$\sigma$ upper limits are indicated for non-detections. The uncertainties do not include the $\sim$10--15\,\%\ absolute calibration error. All the wavelengths are rest-frame. $^{(a)}$ The monochromatic 10 and 12\micron\ fluxes are measured in the CC spectra of each region (see Section \ref{ss:image_modeling}). $^{(b)}$ 24.1\micron\ fluxes derived from the CC $Q$-band imaging (see Section \ref{ss:image_modeling}). $^{(c)}$ Flux of the 11.3\micron\ PAH feature. $^{(d)}$ EW of the 11.3\micron\ PAH. $^{(e)}$ Pa$\alpha$ flux measured in the continuum subtracted F190N NICMOS images \citep{AAH06s}. To convert to flux units, these values should be multiplied by 1.56$\times$10$^{-14}$ erg\,cm$^{-2}$\,s$^{-1}$\,mJy$^{-1}$.
\end{table*}

\subsection{Spectral Modeling}\label{ss:modeling}

For the five selected regions, we decomposed the $N$-band spectra together with the 24\micron\ photometry using a two component model consisting of a modified black-body with $\beta=2$ and a PAH emission template. The latter is derived from the \Spitzer\slash IRS starburst template presented by \citet{Smith07} after removing the dust continuum emission (see \citealt{Pereira2015not} for details). We excluded the low-atmospheric transmission spectral ranges marked in Figure \ref{fig:cc_spec} for the fitting.

The black-body temperature and the intensities of the black-body and the PAH template are free parameters of the model.
In addition, we let the relative strength of the 11.3\micron\ PAH feature free during the fit since the strength of the different PAH features varies both in starbursts \citep{Smith07} and Seyfert galaxies \citep{Diamond2010}.

We calculated the warm dust mass using the following relation
\begin{equation}
M_{\rm dust} = \frac{D^2 f_{\nu}}{\kappa_{\nu} B_{\nu} (T_{\rm dust})}
\end{equation}
where $D$ is the distance, $f_{\nu}$ the observed flux, $\kappa_{\nu}$ the absorption opacity coefficient, and $B_{\nu} (T_{\rm dust})$ the Planck's blackbody law, all of them evaluated at 10\micron. We assumed $\kappa_{10\mu m}=1920$\,cm$^{2}$\,g$^{-1}$  \citep{Li2001}.

\begin{figure}
\centering
\includegraphics[width=0.46\textwidth]{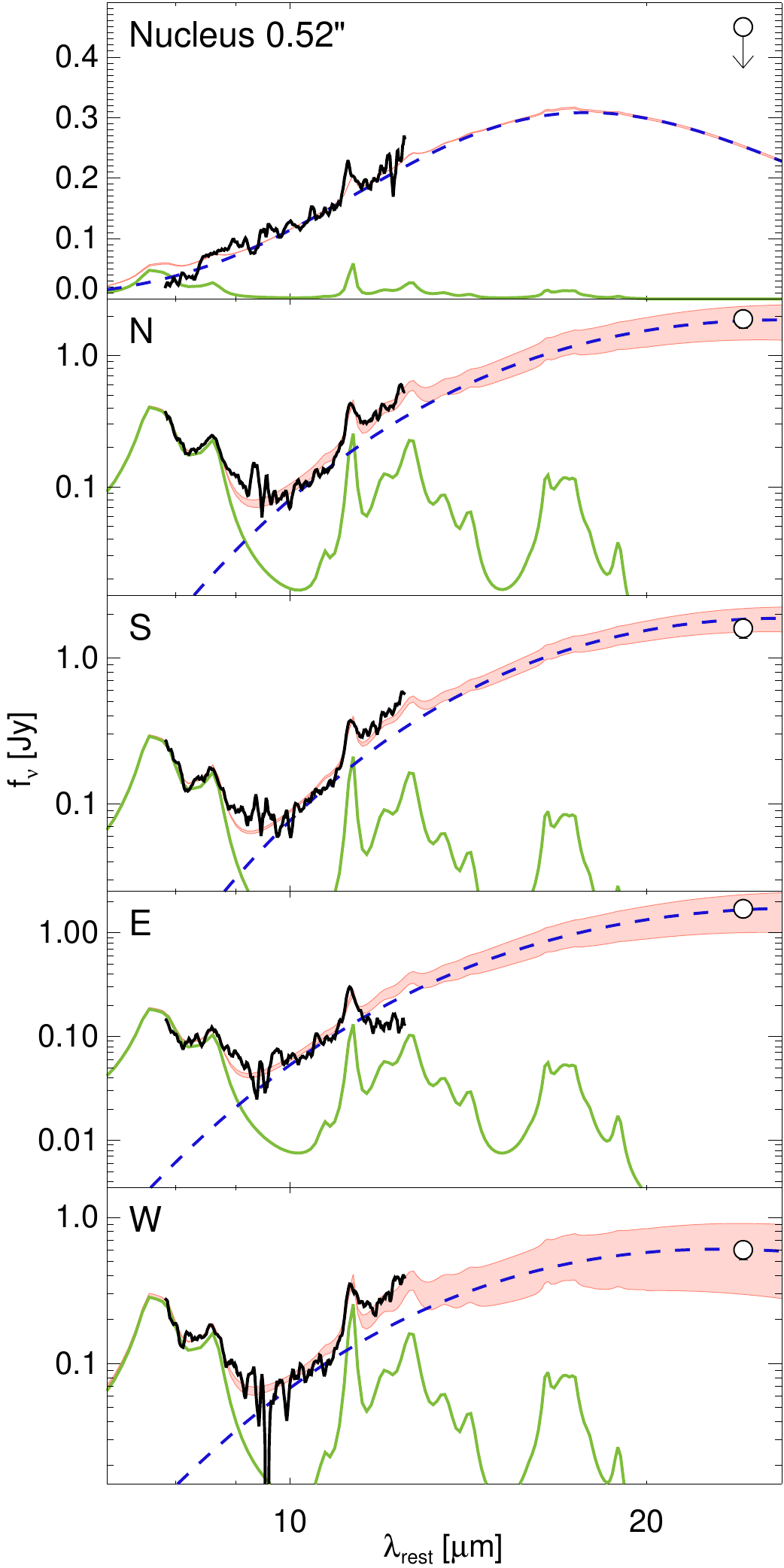}
\caption{Best-fit of the modified blackbody+PAH template models to the CC spectroscopy (solid black line) and 24.5\micron\ photometry (white circle) of the different regions of NGC~1614. The 1$\sigma$ range of the best-fit model is indicated by the red shaded area. The solid green line and the dashed blue line represent the PAH template and the modified blackbody continuum, respectively.
\label{fig:models}}
\end{figure}

The results of the fits are shown in Table \ref{tab:models} and Figure \ref{fig:models}. The mid-IR emission of the SF ring regions are well fitted by a combination of a PAH component, which dominates the emission below 9\micron, and a warm ($T\sim$110\,K) dust continuum component which dominates the emission at longer wavelengths. By contrast, the nuclear 8--13\micron\ spectrum is completely dominated by a warmer ($T\sim$160\,K) dust component.

\begin{table}
\centering
\caption{Results from the modeling of the CC data\label{tab:models}}
\begin{tabular}{@{}lcccccccccc@{}}
\hline
Region & $\frac{\rm 7.7\micron\ PAH}{\rm 11.3\micron\ PAH}$\,$^a$ & $T^{\rm warm}_{\rm dust}$\,$^b$ & $M^{\rm warm}_{\rm dust}$\,$^c$ \\
& & (K) & (10$^3$\Msun)  \\
\hline
Nucleus & 1.7 $\pm$ 0.2 & 160 $\pm$ 8 & 0.2$^{+0.3}_{-0.05}$ \\
N  & 3.6 $\pm$ 0.5 & 110 $\pm$ 3 & 7.6$^{+6.7}_{-3.4}$ \\
S  & 3.1 $\pm$ 0.4 & 109 $\pm$ 4 & 8.9$^{+3.2}_{-2.3}$ \\
E  & 2.9 $\pm$ 0.5 & 108 $\pm$ 6 & 6.3$^{+1.8}_{-4.3}$ \\
W  & 3.3 $\pm$ 0.6 & 115 $\pm$ 3 & 3.9$^{+2.2}_{-1.3}$ \\
\hline
\end{tabular}

\medskip
\raggedright \textbf{Notes:} $^{(a)}$ Ratio between the intensities of the modeled 7.7 and 11.3\micron\ PAH features (Section \ref{ss:modeling}). $^{(b)}$ Temperature of the warm dust component detected in the mid-IR. $^{(c)}$ Mass of the warm dust (see Section \ref{ss:modeling} for details).
\end{table}

\section{The AGN or SF Nature of the Nucleus}\label{ss:agn_vs_sf}

The nature of the nucleus (central 150\,pc) of NGC~1614 is not well established. In part, this is because it is surrounded by a circumnuclear ring with strong star-formation (ring SFR$\sim$40\,\Msun\,yr$^{-1}$; \citealt{AAH01}), which masks the relatively weak nuclear emission when observed at lower angular resolutions.

In the high-angular resolution (0\farcs11) \textit{HST}\slash NICMOS images, the nucleus is slightly resolved and shows near-IR colors compatible with stellar emission, although the CO index is inconsistent with an old stellar population \citep{AAH01}. To explain this, \citet{AAH01} suggested that the nuclear SF is more evolved than that of the star-forming ring.

Based on \textit{ASCA} X-ray observations, \citet{Risaliti2000} suggested that NGC~1614 may host a Compton-thick AGN. However, the Fe\,K 6.4\,keV line, which usually has a high EW in Compton-thick AGNs (although it depends on the obscuring matter geometry; see e.g., \citealt{Fabian2002}), is not detected in more sensitive \textit{XMM-Newton} observations \citep{Pereira2011}.
More recently, the non-detection of CO(6--5) emission and 435\micron\ continuum in the nucleus in high-resolution ALMA observations implies that the amount of dust and molecular gas is much lower than that expected for a Compton-thick AGN \citep{Xu2015}.  Similarly, interferometric radio continuum observations reveal that the nuclear emission is mostly thermal and relatively weak, which also supports the non-AGN nuclear activity \citep{Herrero-Illana2014}. Consequently, SF, as traced by the nuclear Pa$\alpha$ emission, would be the dominant energy source of the nucleus of NGC~1614.

Our new mid-IR data challenge these previous results. The nuclear spectrum shows a strong 12\micron\ mid-IR \hbox{continuum} (and a low EW of the 11.3\micron\ PAH feature), and the nuclear 24\micron\ continuum is weak in comparison with the SF ring emission. Therefore, the nucleus of NGC~1614 presents some characteristics (weak X-ray and far-IR emissions, lacking molecular gas, strong 12\micron\ mid-IR continuum, and Pa$\alpha$ emission) that cannot be explained in a standard AGN or SF context. In the following, we discuss possible modifications to the AGN and SF scenarios to explain the observations available so far.

\subsection{X-ray weak AGN?}

\begin{figure}
\centering
\includegraphics[width=0.45\textwidth]{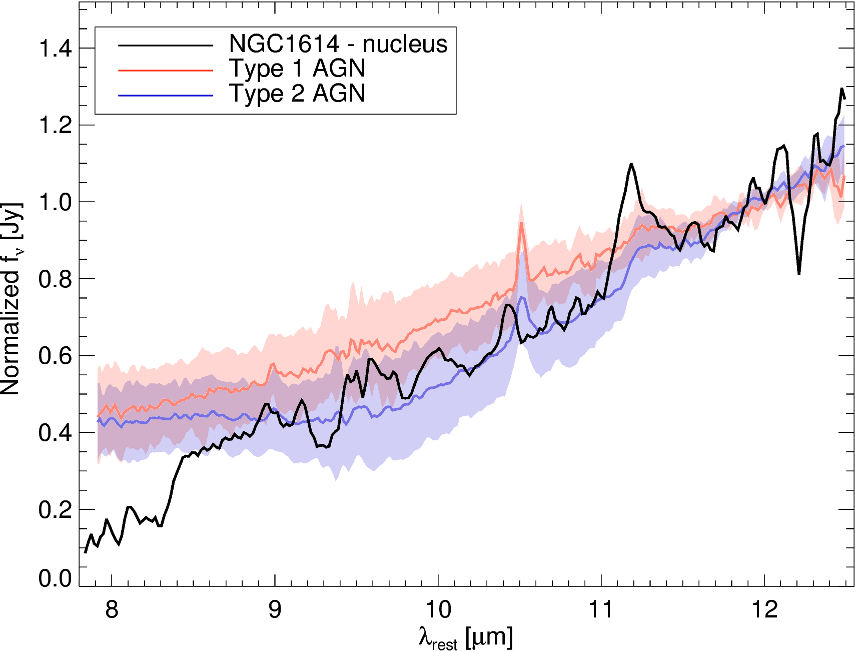}
\caption{Comparison of the nuclear NGC~1614 mid-IR spectrum (black) and the average spectra of Type 1 (red) and Type 2 (blue) Seyfert galaxies from \citet{AAH2014} normalized at 12\micron. The shaded regions represent the 1$\sigma$ dispersion of the averaged spectra.
\label{fig:agn_comparison}}
\end{figure}

\subsubsection{Mid-IR AGN evidences}

The CC spectrum of the nucleus is remarkably different from the spectra of the star-forming regions in the ring of SF. It shows a strong mid-IR continuum relative to the PAH emission, a dust temperature higher than in the SF regions of the ring (Table \ref{tab:models}), and a continuum peak at around $\sim$20\micron\ (Figure \ref{fig:models}). 

Differences in the dust continuum emission are also evident if we consider the 24\micron\ to 10\micron\ flux ratio which is $\sim$10-30 in the ring and $<$5 in the nucleus (see Table \ref{tab:fluxes} and Figure \ref{fig:maps}).
Low 24\micron\slash 10\micron\ ratios are predicted by AGN torus models because of the high dust temperatures reached in the torus (150--1500\,K; \citealt{Nenkova2008}). Moreover, mid-IR \Spitzer\slash IRS spectroscopy of active galaxies shows that for $\sim$30\,\%\ of them (including Type 1 and 2 Seyfert objects) the mid-IR spectra peaks at $\sim$20\micron\ indicating that a warm dust component ($T\sim150-170$\,K) dominates the mid-IR emission \citep{Buchanan2006, Wu2009}. This trend is also observed in ground-based sub-arcsecond mid-IR spectroscopic surveys of Seyfert galaxies (e.g., \citealt{AAH2011torus, RamosAlmeida2009, RamosAlmeida2011}).
In Figure \ref{fig:agn_comparison} we compare the average Type 1 and Type 2 AGN mid-IR spectra obtained by \citet{AAH2014} for nearby Seyfert galaxies. It shows that they all have similar continuum slopes. This suggests that the warm dust conditions in the nucleus of NGC~1614 are similar to those found in Seyfert galaxies.  {Although, the emission of hotter dust (at $\sim$8\micron) is weaker in NGC~1614.}

The minimum 11.3\micron\ PAH EW is located at the nucleus (79 $\pm$ 8)$\times10^{-3}$\micron\ (Table \ref{tab:fluxes}). This behavior is also observed in local Seyfert galaxies, and it is explained in these objects by the increased AGN continuum contribution in the nucleus \citep{AAH2014, Esquej2014, RamosAlmeida2014, GarciaBernete2015}.
In addition, in the nucleus of NGC~1614, the 11.3\micron\ PAH feature is enhanced by a factor of $\sim$2 with respect to the 7.7\micron\ PAH feature (Table \ref{tab:models}). 
Similar enhancements of the 11.3\micron\ PAH feature are observed in active galaxies although on kpc scales \citep{Diamond2010}. 

\subsubsection{Weak X-ray Emission}\label{ss:xrayweak}

A correlation between the 12\micron\ and the 2--10\,keV luminosities is observed for Seyfert galaxies \citep{Horst2008, Levenson2009, Gandhi2009, Asmus2011}. For the nuclear 12\micron\ luminosity measured from the spectrum of NGC~1614 ($\nu L_{\nu}=$2.6$\times$10$^{43}$\,erg\,s$^{-1}$) the expected hard X-ray luminosity would be 1.6$\times$10$^{43}$\,erg\,s$^{-1}$ according to the \citet{Gandhi2009} relation. Threfore, both the nuclear 12\micron\ and expected 2--10\,keV luminosities are comparable to that of an average local Seyfert galaxy (see Figure 1 of \citealt{Gandhi2009}).
However, the observed integrated hard X-ray luminosity of this galaxy is just 1.4$\times$10$^{41}$\,erg\,s$^{-1}$, almost a factor of 200 lower than expected for an AGN, and most of it can be explained by the hard X-ray emission from star-formation (i.e., high-mass X-ray binaries; \citealt{Pereira2011}). Similarly, the soft X-ray emission is also better explained by star-formation \citep{Pereira2011,Herrero-Illana2014}.

If an AGN is present in the nucleus of NGC~1614, three possibilities may explain the weakness of the X-ray emission: it may be a strongly variable source observed during its low state; it may be a Compton-thick AGN so the 2--10\,keV emission is absorbed; or it may be an intrinsically X-ray weak AGN. There are three hard X-ray observations of NGC~1614 during 18\,yr (Table \ref{tab:xray}) which show that the variability is less than a factor of 2. So it is not likely that X-ray variability is the reason for the X-ray weakness. The Compton-thick AGN possibility was rejected by \citet{Xu2015} based on the low amount of molecular gas and cold dust in the nucleus.
Moreover, NGC~1614 is not detected in the 14--195\,keV {\it Swift}\slash BAT 70-Month Hard X-ray Survey \citep{Baumgartner2013}. If NGC~1614 would be a Compton-thick AGN with an intrinsic 2--10\,keV luminosity of 1.6$\times$10$^{43}$\,erg\,s$^{-1}$ (see above), its 14--195\,keV flux would be\footnote{Using \textsc{XSPEC} \citep{Arnaud1996} and assuming a power-law spectrum with $\Gamma=1.9$ \citep{Marconi2004} and $N_{\rm H}=$3$\times$10$^{24}$\,cm$^{-2}$. Increasing the $N_{\rm H}$ up to 10$^{26}$\,cm$^{-2}$ the flux would be reduced by factor of two.} 6$\times$10$^{-11}$\,erg\,cm$^{-2}$\,s$^{-1}$, which is $\sim$4 times the 5$\sigma$ sensitivity of the {\it Swift}\slash BAT survey. 
Finally, it is also possible that the X-ray emission of the NGC~1614 AGN is intrinsically weak. The ultra-luminous IR galaxy Mrk~231 \citep{Teng2014}, as well as several quasars \citep{Leighly2007, Miniutti2012, Luo2014}, have X-ray luminosities 30--100 times weaker than those predicted by the $\alpha_{\rm OX}$\footnote{$\alpha_{\rm OX}=-0.384\log (L(2{\rm \,keV}) \slash L(2500\AA) $ } vs. $L_{\rm 2500A})$ correlation, probably due to a distortion of the accretion disk corona \citep{Miniutti2012, Luo2013}.
In the case of NGC~1614, the nuclear UV emission is completely obscured (see \citealt{Petty2014}), so a direct comparison with the results for these X-ray weak AGNs is not possible. However, using the 12\micron\ emission we obtain that the {observed} 2--10\,keV emission is more than two orders of magnitude lower than the expected value, similar to the X-ray weakness observed on those objects.

\begin{table}
\centering
\caption{2--10\,keV X-ray observations of NGC~1614\label{tab:xray}}
\begin{tabular}{@{}lcccccccccc@{}}
\hline
Date & Telescope & Flux & Ref. \\
& & (10$^{-13}$\,erg\,cm$^{-2}$\,s$^{-1}$)  \\
\hline
1994-02-16 & {\it ASCA} & 5.6 & 1 \\
2003-02-13 & {\it XMM-Newton} & 2.7$\pm$0.4 & 2 \\
2012-04-10 & {\it Swift} & 2.5$\pm$0.4 & 3 \\
\hline
\end{tabular}

\medskip
\raggedright \textbf{References:} (1) \citealt{Risaliti2000}; (2) \citealt{Pereira2011}; (3) \citealt{Evans2014}.
\end{table}

\subsection{or Nuclear Star-formation?}

Alternatively, it is possible to explain the nuclear observations assuming only star-formation (SF). However, the nuclear SF and the SF taking place in the ring surrounding the nucleus must have very different characteristics. In particular, the nuclear mid-IR spectrum shows a strong 8--12\micron\ continuum that is not present in the ring spectra (Figure \ref{fig:models}), and the nucleus remains undetected in the 435\micron\ far-IR continuum and CO(3--2) maps \citep{Xu2015, Usero2015} while the ring is clearly detected.

In our nuclear mid-IR spectrum, we detect the 11.3\micron\ PAH feature which is usually associated with SF (mostly B stars, see \citealt{Peeters2004}). 
Using the $L_{\rm 11.3\mu m\,PAH}$ SFR calibration of \citet{Diamond-Stanic2012}, we estimate a nuclear SFR of $\sim$0.9\,\Msun\,yr$^{-1}$ (Table \ref{tab:sfr} and see Section \ref{ss:sfr_tracers}).
We also used the nuclear Pa$\alpha$ flux \citep{DiazSantos2008} to derive a SFR $\sim$1.5\,\Msun\,yr$^{-1}$ (assuming $A_{\rm k}=$0.3\,mag; \citealt{AAH01}), so both SFR tracers are in agreement within a factor of 2.
Finally, we used the IR continuum upper limits at 24 and 432\micron\ to derive an upper limit for the nuclear IR (4-1000\micron) luminosity of $<$6$\times$10$^{43}$\,erg\,s$^{-1}$. This upper limit is compatible with the expected IR luminosity for a SFR$\sim$1.5\,\Msun\,yr$^{-1}$ ($\sim$4$\times$10$^{43}$\,erg\,s$^{-1}$; \citealt{Kennicutt2012}). Therefore, all these IR SFR tracers are compatible and they indicate that the nuclear SFR is $\leq$1.5\,\Msun\,yr$^{-1}$, that is, less than $<$2\,\%\ of the total SFR of NGC~1614 ($\sim$100\,\Msun\,yr$^{-1}$; \citealt{Pereira2015not}).

However, the nuclear and the integrated IR (8--500\micron) spectral energy distributions are very different. The ring is detected at 435\micron\ \citep{Xu2015} and 24\micron\ (Figure \ref{fig:maps}),but the nucleus is not.
Therefore, this implies that the dust temperature is much higher in the nucleus, as already suggested by our mid-IR data. This higher nuclear dust temperature (Table \ref{tab:models}) can be explained by the enhanced radiation field density, which is expected to increase the dust temperature (see \citealt{Draine07}), due to an increased density of young stars in the nucleus (or an AGN, see Section \ref{ss:xrayweak}).

Molecular gas is not detected in the nucleus of NGC~1614. From the 0\farcs5 resolution CO(3-2) ALMA observations of NGC~1614, \citet{Usero2015} estimate an upper limit to the nuclear molecular gas mass of 3$\times$10$^{6}$\,\Msun\footnote{Assuming a CO(3-2) to CO(1-0) ratio of $\sim$1 and the Galactic CO-to-H$_2$ conversion factor \citep{Bolatto2013}. Using the conversion factor for ULIRGs it would be a factor of $\sim$4 lower.}. This low molecular gas mass puts the nucleus of NGC~1614 well above the Kennicutt-Schmidt relation (see Figure 8 of \citealt{Xu2015}).
Consequently, the molecular gas depletion time is $<$3\,Myr, much lower than in normal galaxies at 100\,pc scales (1--3\,Gyr; e.g., \citealt{Leroy2013}), and also lower than in local ULIRGs (70--100\,Myr; e.g., \citealt{Combes2013}).
A short depletion time might indicate that the ignition of the nuclear SF occurred earlier than in the ring (see \citealt{AAH01}). {Therefore, the nuclear starburst would have consumed a larger fraction of the original cold molecular gas than the younger starburst of the ring. Actually, the evolutionary state of the SF regions is commonly used to explain the dispersion of individual SF regions in the Kennicutt-Schmidt relation (e.g., \citealt{Onodera2010, Schruba2010, Kruijssen2014}).} 
However, the integrated (including nucleus and SF ring) dense molecular gas depletion time in NGC~1614 is also shorter ($\sim$10\,Myr) than in other LIRGs ($\sim$50\,Myr; \citealt{GarciaBurillo2012}), so it is not obvious to associate the particularly short nuclear depletion time with older SF.
Alternatively, a massive molecular outflow, produced by an AGN or SN explosions (see \citealt{GarciaBurillo2015}), could have swept most the molecular gas away from the nucleus. 

On the other hand, the hard X-ray luminosity of this object is also compatible with a SF origin \citep{Pereira2011}, although most of the emission would be produced in the ring. Unfortunately, the angular resolution of the \textit{Chandra} X-ray data is not sufficient to separate the nucleus and the ring \citep{Herrero-Illana2014}.

Note that, in principle, a combination of SF and a normal AGN would be also possible. However, this assumption suffers the same problems explaining the observations than the SF and AGN individually. For these reason, we do not discuss this AGN$+$SF composite possibility.

\begin{table}
\centering
\caption{SFR from different IR tracers\label{tab:sfr}}
\begin{tabular}{@{}lcccccccccc@{}}
\hline
& & \multicolumn{3}{c}{SFR ($M_\odot$\,yr$^{-1}$)} \\[-1.5ex]
& \cline{2-4} \\[-2.5ex]
Region & $A_{\rm k}$\,$^{a}$ & Pa$\alpha$\,$^{b}$ & 11.3\micron\ PAH\,$^{c}$ & 24\micron\,$^{d}$ \\
\hline
Nucleus$^\star$ & 0.3 & 1.5 & 0.9 & $<$6 \\
N  & 0.7 & 9.3 &  4.1 &  22 \\
S  & 0.8 & 11.2 & 3.1 &  21 \\
E  & 0.6 & 7.8 &  4.1 &  26 \\
W  & 1.0 & 16.4 & 5.2 &  7 \\
\hline
\end{tabular}

\medskip
\raggedright \textbf{Notes:} $^{(a)}$ $K$-band extinction in magnitudes derived from the stellar colors \citep{AAH01}. 
$^{(b)}$ Extinction corrected Pa$\alpha$ SFR using the \citet{Kennicutt2012} calibration assuming H$\alpha$\slash Pa$\alpha$ = 8.51. 
$^{(c)}$ SFR obtained from the 11.3\micron\ PAH luminosities (Table \ref{tab:fluxes}) based on the \citet{Diamond-Stanic2012} calibration. 
We multiplied by 2 our 11.3\micron\ PAH luminosities to account for the different method used to measure the PAH features (local continuum vs. full decomposition, see \citealt{Smith07}).
$^{(d)}$ SFR derived from the monochromatic 24\micron\ luminosities (Table \ref{tab:fluxes}) using the \citet{Rieke2009} calibration. $^{(\star)}$ Nuclear SFR derived assuming that all the nuclear emission is produced by SFR (i.e., no AGN).
\end{table}

\section{SFR Tracers at $\sim$150\,\lowercase{pc} Scales}\label{ss:sfr_tracers}

Using the new CC mid-IR data (11.3\micron\ PAH and 24\micron\ continuum) in combination with the NICMOS Pa$\alpha$ image, we can test several SFR calibrations at 150\,pc scales in this galaxy.

In Table \ref{tab:sfr} we show a summary of the SFR derived using these tracers for the five regions we defined in NGC~1614. We used the calibrations of \citet{Kennicutt2012}, \citet{Diamond-Stanic2012}, and \citet{Rieke2009} for the Pa$\alpha$, 11.3\micron\ PAH, and 24\micron\ tracers, respectively. The Pa$\alpha$ emission was corrected for extinction using the near-IR continuum colors (see \citealt{AAH01}). Since the extinction corrected Pa$\alpha$ calibration is a direct measurement of the number of ionizing photons produced by young stars, we consider it as the reference SFR tracer.

The 24\micron\ luminosity gives the highest SFR values (2--3 and 5--7 times higher than those derived from the Pa$\alpha$ and 11.3\micron\ PAH luminosities, respectively), except in the W region of the ring. The modeling of the radio emission of the W region indicates the presence of supernovae (SN; \citealt{Herrero-Illana2014}), so it could be more evolved than the rest of the ring. Therefore, a lower amount of young stars would be dust embedded in this region reducing the warm dust emission.

The disagreement between the extinction corrected Pa$\alpha$ and the 24\micron\ SFR values is $\sim$0.4\,dex, which is higher than the calibration uncertainty (0.2\,dex). Although, in principle, both tracers should produce similar SFR estimates (see Equations 5 and 8 of \citealt{Rieke2009}).
There are two possibilities to explain this. First, it is possible that even the extinction corrected Pa$\alpha$ emission underestimates the SFR. In extremely obscured regions (e.g., $A_{\rm v}>$15--20\,mag), dust might absorb the Pa$\alpha$ emission completely, as well as part of the ionizing photons, and therefore, rendering any extinction correction ineffective. Alternatively, an increase of the dust temperature at high SFR densities, like in the SF ring of NGC~1614, can produce enhanced 24\micron\ emission that might not be taken into account by the 24\micron\ SFR calibration which is valid for integrated emission of galaxies (e.g., \citealt{Calzetti2010}). The stellar $A_{\rm k}$ measured in the SF ring of NGC~1614 is 0.6--1.0\,mag ($A_{\rm v}=$\,5--10\,mag; \citealt{AAH01}), so the obscuration level is not as extreme as observed in some ULIRGs ($A_{\rm v}=$\,8--80\,mag; \citealt{Armus07}). In addition, the 9.7\micron\ silicate absorption in the SF ring spectra is not very deep (Figure \ref{fig:models}).
Therefore, this favors the second possibility. That is, an increased 24\micron\ emission in the SF ring of NGC~1614 due to a warmer dust emission.

According to Table \ref{tab:sfr}, the SFR derived from the 11.3\micron\ PAH luminosity is 2--4 times lower than that derived from Pa$\alpha$. The 11.3\micron\ PAH SFR calibration is based on $\sim$kpc integrated measurements \citep{Diamond-Stanic2012}. However, it is known that the PAH emission, and in particular the 11.3\micron\ PAH emission, is more extended than the warm dust continuum and other ionized gas tracers (e.g., [\ion{Ne}{ii}]12.81\micron; \citealt{DiazSantos2011}). Actually, $\sim$30--40\,\%\ of the total PAH emission is not related to recent SF \citep{Crocker2013}. Therefore, this SFR calibration possibly includes a considerable amount of PAH emission not produced by young stars. {In addition, using templates of SF galaxies, \citet{Rieke2009} showed that the 11.3\micron\ PAH contribution to the total IR luminosity drops by a factor of $\sim$2.5 for galaxies with $L_{\rm IR}>10^{11}$\,\Lsun. A similar result was found by \citet{AAH2013} for a sample of local LIRGs.}
{A combination of these reasons might} explain why we obtain these relatively low SFR estimates from the 11.3\micron\ PAH luminosities for the  $\sim$150\,pc SF regions in the ring of NGC~1614.

\section{Conclusions}\label{s:conclusions}

We analyzed new GTC\slash CC high-angular resolution ($\sim$0\farcs5) mid-IR observations of the local LIRG NGC~1614. The new $N$-band spectroscopy and $Q$-band imaging are combined with existing \textit{HST}\slash NICMOS Pa$\alpha$ and T-ReCS 8.7\micron\ images to study the properties of the bright circumnuclear SF ring and the nucleus of this object. The main results are the following:

\begin{enumerate}
\item We extracted mid-IR spectra from four different regions in the circumnuclear SF ring and from the nuclear region (central 0\farcs5$\sim$150\,pc). The spectra from the SF ring are typical of a SF region with strong PAH emission and a shallow 9.7\micron\ silicate absorption. By contrast, the nuclear spectrum has a strong mid-IR continuum, which dominates its mid-IR emission, and weak PAH emission (EW$_{\rm 11.3\micron}$=80$\times$10$^{-3}$\micron). Similarly, the SF ring is clearly detected in the 24.5\micron\ image, as expected for a SF region, while the nucleus is {weaker} at this wavelength.
\item A two component model, consisting of a modified black-body with $\beta=2$ and a PAH emission template, reproduces the observed $N$ spectra and $Q$ photometry well. The main differences between the nuclear and the SF ring observations are: the higher dust temperature in the nucleus (160\,K in the nucleus vs. $\sim$110\,K in the ring); the lower PAH EW; and the lower nuclear 7.7\micron\slash 11.3\micron\ PAH ratio. 
\item The above results based on the mid-IR data, suggest that an AGN might be present in the nucleus. However, this is at odds with the low X-ray luminosity of NGC~1614 ($\sim$200 times lower than that expected for an AGN with the observed 12\micron\ continuum luminosity).
Since the hard (2--10\,keV) X-ray emission shows no variability, and likely it is not a Compton-thick AGN, if an AGN is present in NGC~1614, it must be an intrinsically X-ray weak AGN. We also calculated an upper limit to the IR luminosity of the nucleus, $<$6$\times$10$^{43}$\,erg\,s$^{-1}$.
\item Alternatively, SF can explain the observations of the nucleus too. However, we need to invoke extremely short molecular gas depletion times ($<$3\,Myr {for a nuclear SFR of $\sim$1--1.5\,\Msun\,yr$^{-1}$}), and an increased radiation field density to explain the observed hot dust in the nucleus.
\item Finally, we compared three SFR tracers at 150\,pc scales {in the circumnuclear ring}: extinction corrected Pa$\alpha$, 11.3\micron\ PAH, and 24\micron\ continuum. Since the extinction is not extremely high ($A_{\rm v}<10$\,mag), we take as reference the Pa$\alpha$ derived SFR. In general, the 24\micron\ SFR overestimates the SFR by a factor of 2--3, while the 11.3\micron\ PAH underestimates the SFR by a factor of 2--4. The former might be explained if the dust temperature is higher in the SF regions of NGC~1614, while the latter {could be because we do not include diffuse PAH emission in our measurements as well as because the PAH contribution to the total IR luminosity might be reduced in LIRGs.}
\item In the West region of the ring, the 24\micron\ emission is $\sim$5 times weaker than expected based on the observed Pa$\alpha$\slash 24\micron\ ratio in this galaxy. We propose that this is because this is a more evolved SF region (SN are present; \citealt{Herrero-Illana2014}) where a larger fraction of the young stars are not dust embedded.
\end{enumerate}

In summary, our mid-IR data suggest that an intrinsically X-ray weak AGN ($L^{\rm AGN}_{\rm bol}\sim$10$^{43}$\,erg\,s$^{-1}$, $<$5\,\%\ of the NGC~1614 bolometric luminosity) might be present in the nucleus of NGC~1614. However, SF with a short molecular gas depletion time and increased dust temperatures can explain the observations as well. In order to further investigate the nature of the nucleus of this galaxy, IR and sub-mm high-angular resolution observations are needed.

\section*{Acknowledgments}

We thank the anonymous referee for useful comments and suggestions.
We thank the GTC staff for their continued support on the CanariCam observations.
We acknowledge support from the Spanish Plan Nacional de Astronom\'ia y Astrof\'isica through grants AYA2010-21161-C02-01, and AYA2012-32295.
AAH and AA acknowledges funding from the Spanish Ministry of Economy and Competitiveness under grants AYA2012-31447 and AYA2012-38491-CO2-02, which are party funded by the FEDER program.
MAPT acknowledges support from the Spanish MICINN through grant AYA2012-38491-C02-02. 
CRA acknowledges support from a Marie Curie Intra European Fellowship within the 7th European Community
Framework Programme (PIEF-GA-2012-327934).
Based on observations made with the Gran Telescopio Canarias (GTC), installed in the Spanish Observatorio del Roque de los Muchachos of the Instituto de Astrof\'isica de Canarias, in the island of La Palma.
Partially based on observations obtained at the Gemini Observatory (program GS-2006B-Q-9), which is operated by the Association of Universities for Research in Astronomy, Inc., under a cooperative agreement with the NSF on behalf of the Gemini partnership: the National Science Foundation (United States), the National Research Council (Canada), CONICYT (Chile), the Australian Research Council (Australia), Minist\'erio da Ci\^encia, Tecnologia e
Inova\c{c}\~ao (Brazil) and Ministerio de Ciencia, Tecnolog\'ia e Innovaci\'on Productiva (Argentina).

\label{lastpage}

\end{document}